# Conservation Laws for Crystal of Topological Defects

**Author and Affiliation**

Jieh-Wen Tsung*
Electrophysics Department, National Chiao-Tung University, Hsinchu, Taiwan

**Abstract**

Stable array of point defects was generated in nematic liquid crystal. Point defects with topological charge of +1 and -1 (hedgehogs) were generated in vertically aligned liquid crystal cell. The hedgehogs were arranged in square or hexagonal arrays, and the shape of defect, such as radial or circular hedgehogs, were under the control of delicately patterned electrodes. Base on the two-dimensional defect array discovered in experiments, three-dimensional crystals of defects were postulated. The flux of the liquid crystal director field was summarized for each unit cell. The analysis showed that the crystal structure is periodic and stable if the total flux in a unit cell is zero. The surface and bulk topological charges of each unit cells were analyzed. The summation of topological charges of a unit cell obeys the topological rules of a homeotropic droplet. The hidden conservation laws of flux and topological charges in a "soft crystal of defects" were discovered, which can be implemented in templates for self-assembled micro-structure and tissue design. Most important of all, the conservation laws may answer why natural structures and tissues grow in to a particular shape and form.

**Key Words**

Topological defect, symmetry breaking, liquid crystal

*Corresponding Author
e-mail: jwtsung@nctu.edu.tw

**Introduction**

Topological defect arises when the symmetry of order in material is broken. The symmetry breaking can be induced by phase transitions or by application of confinements. The types of defects, such as point, line, or walls, depends on dimension of the order[1,2] and topology of the confinement[3,4]. Though topological defects represent the corruption of the previous order, they trap molecules[5] and particles[6,7]. The network between defects provide a scaffold[7–9], or the attraction between defects[10,11] triggers the self-assembly and holds the inclusions together[12,13]. Periodic, regular, stable topological defect network mediates the birth of a new order. The dynamics and mechanism of defect network formation explain why and how colloids[10–13], biological fibers[5,14–16], and tissues[9,17,18] grow into a particular structure.

Two-dimensional (2D) colloidal crystal assembled by attraction between topological defects in nematic liquid crystal (NLC) was discovered in 2006[19]. The "boojum" point defect attached on the particle generates a dipolar director field. The positive and negative poles of the dipoles attract each other, and the particles line up in a chain. The chains attract each other when the dipoles are anti-parallel, forming a dipolar crystal by self-assembly. The "Saturn ring" loop defect around the waist of the particle generates a quadrupolar director field, resulting in the hexagonal crystals. Three-dimensional (3D) dipolar crystal was successfully assembled in 2013[13]. Since 2015, pure topological defect arrays without inclusions have been generated by using crossed strip electrodes[20], patterned surface alignment[7,21], micro cavities on wafer[22], doping of ions in NLC[23], standing pressure wave[24], and interaction between optical vortex and liquid crystal[25]. These methods set up periodic confinements, and the symmetry breaking of NLC in periodic confinements leads to array of umbilical escapes or disclinations. The defect arrays are topologically protected and self-retained. Therefore, they are promising templates for wires[8], fibers, or tissues.

In analogy to mineral crystals, defect array has periodic pattern that exhibit long-range order[26–28]. One topological defect (or a group of topological defects) can be one unit cell, and the periodic repetition gives a crystal. In mineral crystals, the electron configuration of atoms determines the rotational symmetry of the unit cell, and the periodic repetition of unit cells renders the rotational symmetry of the crystal. In NLC, the symmetry of isotropic distribution is broken by the rod-like shape of molecules, leading to long-range nematic order. Then the symmetry of nematic order is broken by periodic confinements again, leading to topological defect crystals having "extra" long-range order. There are two hierarchies of orders: the nematic order and the periodic arrangement of defects. The shape of the topological defect, such as radial, circular, hyperbolic hedgehogs, lines, walls, which are associated with topological charges, must match to the periodicity of the defect crystal. The director field across the boundary must be continuous, except the locations of singularities. The strain between the defect must have static balance, so the crystal of topological defects can be stable. Based on the above mentioned principles, selection rules of stable topological defect array can be developed. Above the selection rules, conservation laws of director field may be hidden in the structure of topological defect array.

To observe and analyze the defect crystal systematically, topological defects were generated in pixelized NLC test cell and arranged on square or hexagonal arrays. For the first time, large arrays of points defect with designed shapes were created. The distinction between stable and unstable crystals were clear. Based on the topology and symmetries of the director field, conservation laws of director flux and topological charges were discovered.

## Methods

### Structure of the liquid crystal cell

Array of topological defects was generated in test cells with homeotropic alignment on the surface. NLC with negative dielectric anisotropy ($\Delta\varepsilon$) of –3.0 was sandwiched between glass substrates with Indium-Tin-Oxide (ITO) sputtered on the surface. The patterns of ITO were made of lithography and wet etching. Polyimide (PI) was coated on the glass substrate, so the NLC stood vertically on the surfaces. The bottom substrate was patterned, and the top substrates was a full ITO (see Fig. 1(a)). NLC was injected into the test cell by capillary effects. Square wave AC voltage with frequency of 60 Hz was applied on the bottom ITO, and top ITO was connected to the ground.

The pixel array was patterned on the bottom substrate (Fig. 1(a)). The pattern of individual pixels could be a pad, a fishbone, or a coil (Fig. 1(b)). Then the individual pixels were arranged in square or hexagonal lattices (Fig. 1(c)). The topological defects generated by the pixels were be arranged as if a 2D crystal, and each pixel was a unit cell.

When the voltage is applied, the ordinary axis of negative NLC align along the direction of the electric field, which means that the vertically standing NLC will recline and lie in plane (the x-y plane in Fig. 2(a)). In a pixel as shown in Fig. 2(a), NLC on the edge of the bottom electrode reclines and points into the center of the pixel due to strong fringe field. The vertically aligned NLC on the "celling (top)", on the "floor (bottom)", and the reclined NLC on the "walls" compose a "box" with homeotropic surfaces. The pixel becomes a closed homeotropic boundary when electric field is applied. A homeotropic box is topologically equivalent to a homeotropic sphere, in which the liquid crystal configuration can be an umbilical escape (ubmilic)[29] or a point defect[3].

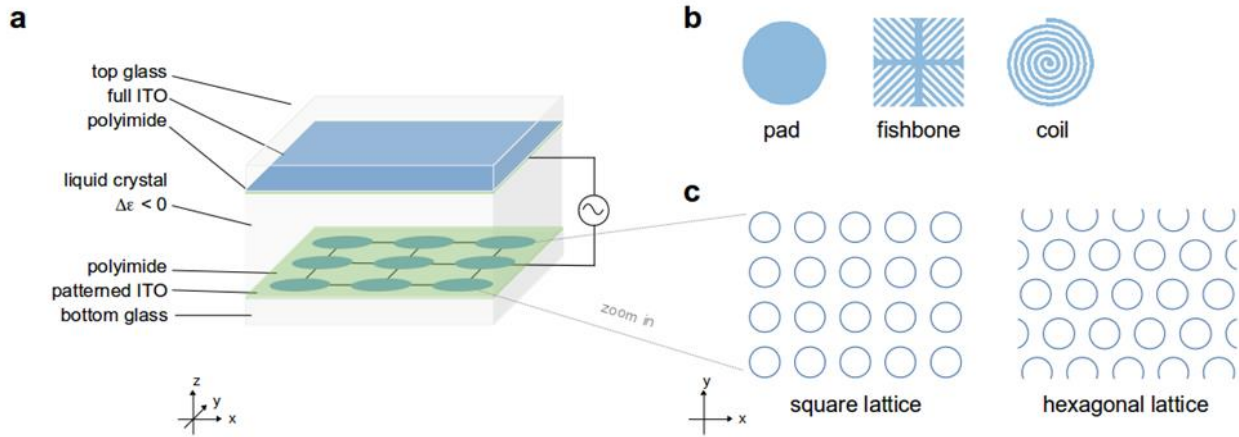

**Figure 1 Structure of the liquid crystal cell.** (a) 3D schematic diagram of the test cell. The bottom Indium-Tin-Oxide (ITO) is a pixel array. The top ITO is full. Liquid crystal is with negative dielectric anisotropy ($\Delta\varepsilon<0$), and aligned vertically on the surface with the polyimide (PI). (b) Candidates of pixels. The pattern can be a pad, a fishbone, or a coil. (c) Array of pixels, arranged in square or hexagonal lattices.

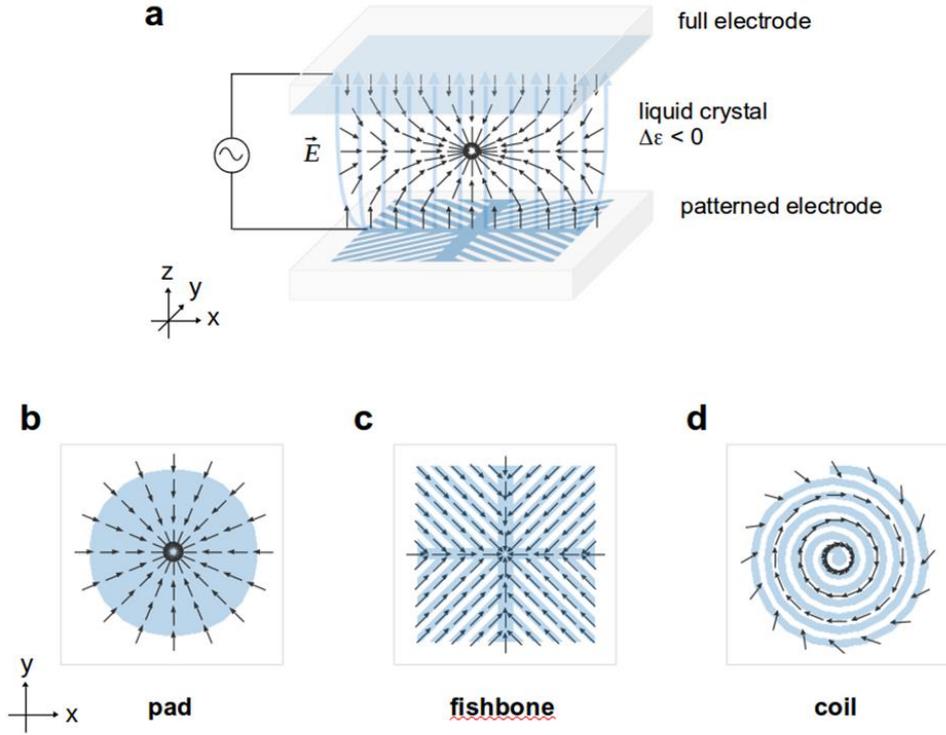

**Figure 2 Director field in a vertically aligned liquid crystal cell.** (a) 3D schematic of a pixel. The light blue, thick arrows indicate the electric field, and the black short arrows show the director of liquid crystal. (b), (c), (d) are top view of the pad, fishbone, and coil pixels, respectively, showing the liquid crystal configuration while electric field is applied.

A pad is the basic shape with Euler characteristic number ($\chi$) equal to 1, which can continuously morph into a fishbone or a coil. Every pixel is a closed box with $\chi$ of 2. Based on the Gauss-Bonnet[30] and Gauss-Stein Theorem[4], the total topological charge of defects in a closed, homeotropic boundary is equal to $\chi/2$. The pad, fishbone, and coil are topologically identical, so the defect generated by each of them must have the same topological charge. NLC tends to align along the strip due to the fringe field[31,32]. Therefore, fishbone and coil lead to radial (Fig. 2(c)) and circular hedgehogs (Fig. 2(d)), respectively. On a pad, the location of the point defect (or umbilics) is arbitrary. On fishbone and coil, the liquid crystal configuration has reflectional and rotational symmetry. Therefore, the point defects (or umbilics) can be fixed at the center of the pixel because of static balance.

In this research, the splay and bend elastic constants of NLC were 15 pN, the cell gap was 3 μm, and the optimum width and spacing of the strips were chosen to be 3 μm[31,32]. The size of a unit cell was 100 μm × 100 μm or 50 μm × 50 μm. The pixels were connected along the axis of transnational symmetry. In square lattice, the connection between pixels were on 0º, 90º, 180º and 270º. In hexagonal lattice, the connection between pixels were on 0º, 60º, 120º, 180º, 240º, and 300º (see Fig. 3 for pictures of pixel array under optical microscope).

The pixel arrays of pad, fishbone, and coil were fabricated by lithography and wet-eching. Arrays were arranged in either square or hexagonal lattices. By using the delicately designed ITO, arrays of point defects (or umbilics) were generated. The director field were analyzed under optical microscope (OM) and polarized optical microscope (POM).

## Results

**Identification of topological defects**

Pictures of bare test cells under OM is shown in Fig. 3. Every pixel is a closed homeotropic boundary with χ equal to 2, so one point defect appears at each pixel (the dark spots in Fig. 3), as expected by the Gauss-Stein Theorem. The dark spot is a clear sign of discontinuity in liquid crystal[33], because light through the defect core exhibit strong diffraction[34]. The destructive interference of diffracted light leads to the dark spot with texture like a hole[35] (I need a reference on experiment). Light through continuous distortions, for example, umbilical escapes, shows lensing effect[36,37] without a diffracting dark spot.

Additional to the point defects at the center of pixels, extra point defects appear between the pixels on the connection bars. In a square lattice, a defect in the pixel is surrounded by four extra defects. Pad, fishbone, and coil pixels result in the same structure. In hexagonal lattice, six extra defect is generated surrounding one pixel. Again, the pattern of an individual pixel does not affect the structure of defect array.

At the middle of the four (three) corners of four neighboring square (hexagonal) pixels, the director field should be a saddle deformation (saddle)[38,39], as illustrated in Fig. 4. When the pixels are very close to each other, or when the electric field is strong, the saddle shrinks, and creates a director field similar to a reversed hyperbolic hedgehog. The saddle appeared as a faint, circular shade in the OM pictures (see the pictures in Fig. 4). Defects and saddles can be clearly distinguished by the texture. Defect is a solid hole, and saddle is just a circular shadow. In pictures taken under POM, defect is surrounded by circular diffraction pattern, while saddles do not show diffractions (refer to Fig. 6(c)). When the focus plane of OM moved from the top substrate to the bottom substrate, the image of defects evolved from concentric rings to a black dot, and then to a blurred, dark spot. This is the typical feature of Fresnel diffraction of light through a pinhole[34,35]. On the other hand, when the focus plane moved from the top to the bottom, the image of saddle changed from a big faint shade to a smaller one, and then suddenly transformed into a faint, bright spot, which is barely recognized. Studies based on ray tracing in liquid crystal lens[36,37] showed that a hyperbolic hedgehog act as a diverging lens. Hence, the dark shade above the saddle is a result of deflected light, and the barley recognized bright spot could be the virtual image of the diverging lens.

Point defects and saddles were arranged in square and hexagonal arrays, like atoms arranged in square and hexagonal crystal lattices. The position of the defects and saddles were fixed by the rotational symmetry in the unit cell and reflectional symmetry between the unit cells.

**Director field of the defect network**

The director field was recognized by placing the liquid crystal cell between crossed polarizes under POM. A λ–wave plate was inserted between the cell and analyzer to add optical retardation on the 135° direction w.r.t. to the polarizer, so the directors on 45° and 135° can be clearly distinguished by the birefringence colors. Here 45° and 135° are in lime and pink, respectively, as shown Fig. 5. Liquid crystal on the edge of the electrode is not completely in plane, leading to blue and orange colors of shorter optical retardation.

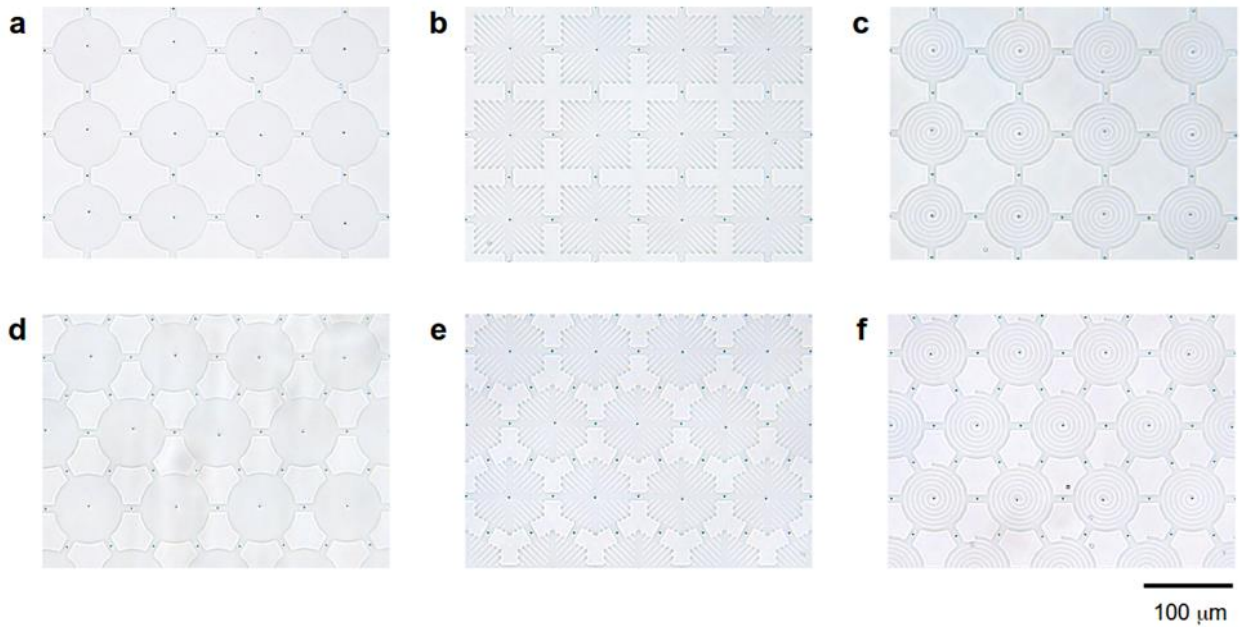

**Figure 3 Array of point defects under optical microscope.** Point defects (the black spots) in liquid crystal were generated by patterned ITO. (a) Pad pixel in square lattice. (b) Fishbone pixel in square lattice. (c) Coil pixels in square lattice. (d) Pad pixel in hexagonal lattice. (e) Fishbone pixel in hexagonal lattice. (f) Coil in hexagonal lattice.

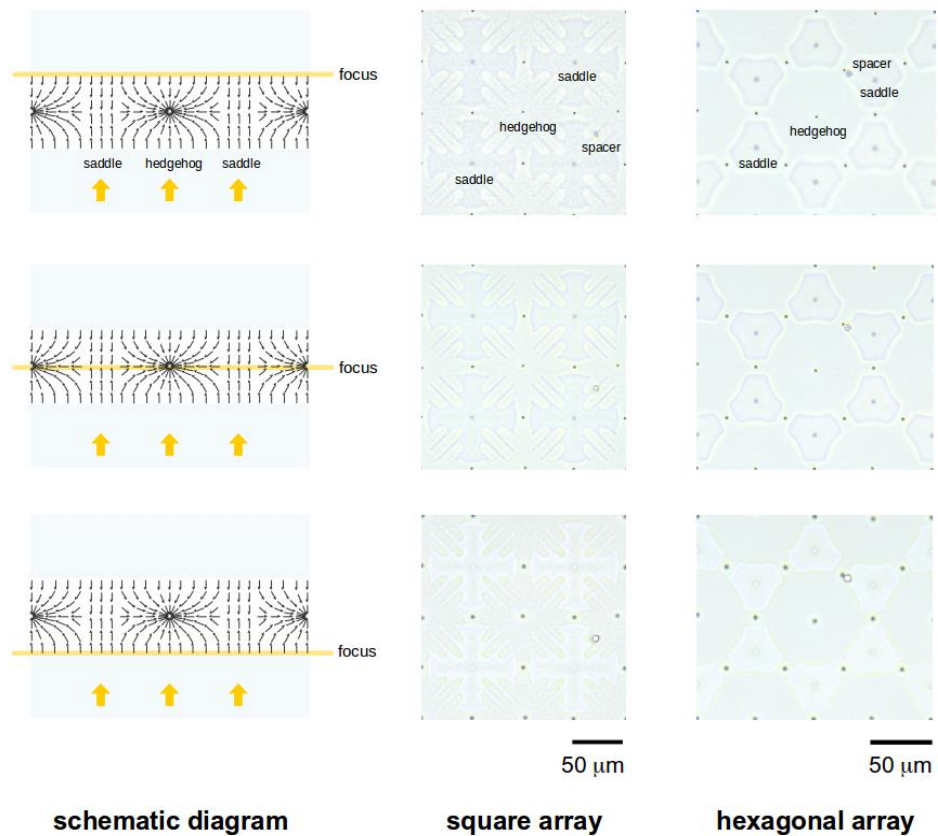

**Figure 4 Optically sectioned images of defect array.** Schematic diagram of director field is on the left. Black arrows show the directors. Yellow thick shades indicate the position of focal planes. Yellow thick arrows indicate the direction of light. Bright field OM pictures of square and hexagonal defect array is on the right. Pictures on the top, middle, and bottom rows were taken when the focus is on the top substrate, on the defect, and on the bottom substrate, respectively. The ball spacer is the indicator for the location of the focus plane.

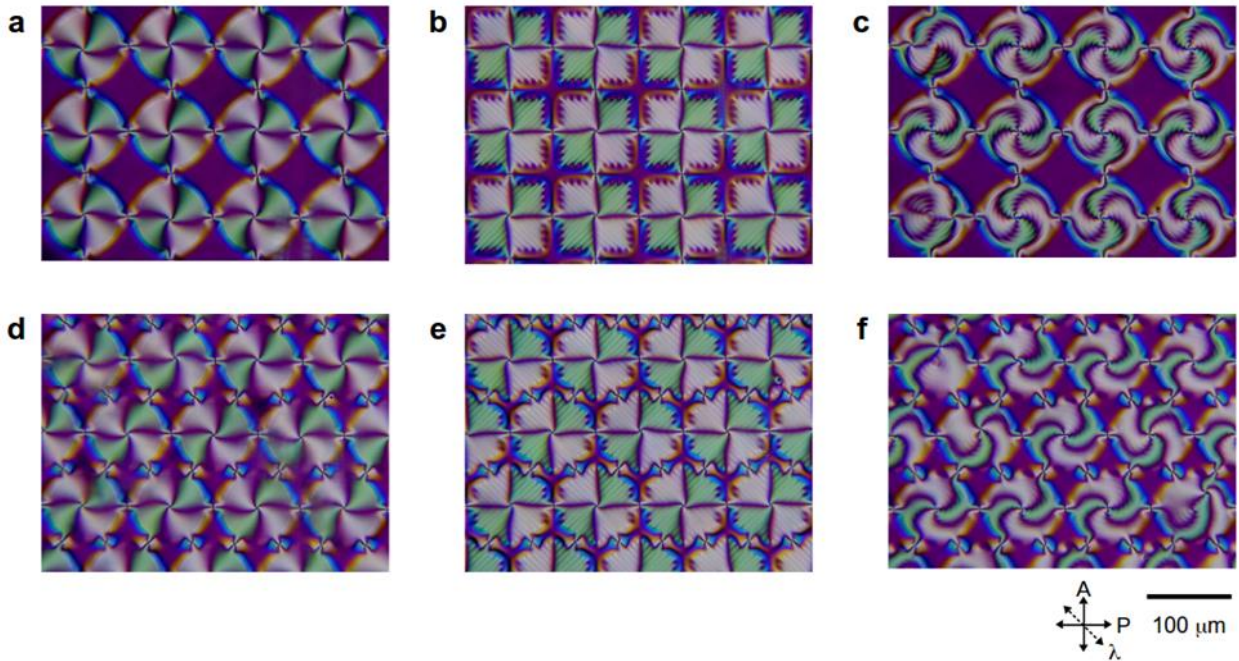

**Figure 5 Array of point defects under polarized optical microscope.** Crossed polarizers on 0° and 90°. A λ-wave plate was placed on 135°, so the directors lying in 45° and 135° show lime and pink, respectively. (a) Pad pixel in square lattice. (b) Fishbone pixel in square lattice. (c) Coil pixels in square lattice. (d) Pad pixel in hexagonal lattice. (e) Fishbone pixel in hexagonal lattice. (f) Coil in hexagonal lattice.

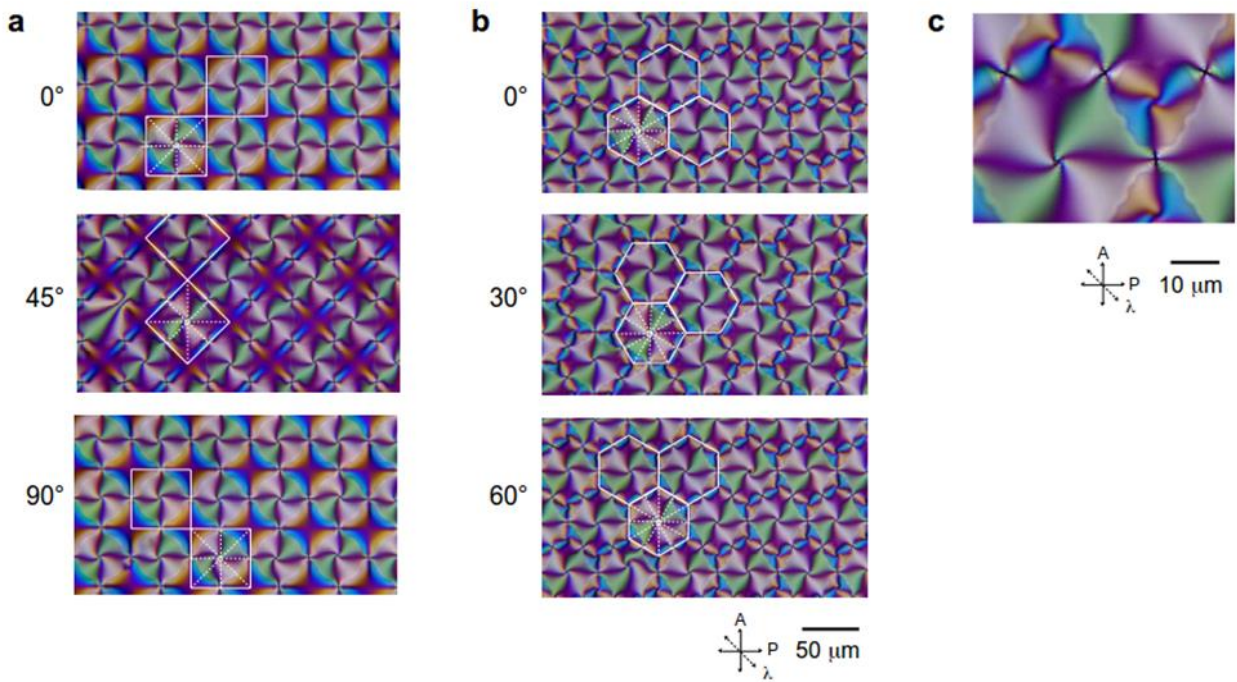

**Figure 6 Rotation of defect array under polarized microscope.** Crossed polarizers on 0° and 90°. λ-wave plate on 135°. (a) Square array lying in 0°, 45° and 90°. (b) Hexagonal array lying in 0°, 30° and 60°. White square and hexagonal box indicates the domain boundary of a unit cell. Dashed, white lines show the simplified director field. (c) Diffraction of light through the point defects.

Topological charge of the defects was identified by Shlieren textures under POM. The topological charge is the number of brushes divided by 4. All the defects had four dark brushes, meaning that the topological charge was either +1 or -1. Sign of defect was examined by rotating the polarizers. All the defects at the center of the pixel rotated at the same direction with the polarizes, but the defects on the connecting bars of pixels rotated in an opposite direction. Therefore, the defect at the pixel was a radial (or spiral) hedgehog, surrounded by four (six) hyperbolic hedgehogs in square (hexagonal) lattices. In 2D space, topological charges of radial and hyperbolic hedgehog are +1 and -1, respectively. The network of hyperbolic hedgehogs is like the scaffold of the 2D crystal. The structure of the scaffold is determined by the shape of the lattice (square or hexagonal), unaffected by the radial (or spiral) hedgehog at the center of the pixel.

The rotational symmetry in the director field was identified by rotating the test cell while the polarizer, analyzer, and wave plate were fixed (Fig. 6). The domain boundaries of a unit cell were recognized by the rotation, too. When the boundary was parallel to the polarizer or the analyzer, it shows a thin, dark line. Every 90° of rotation of a square array gave the same optical image (Fig. 6(a)), meaning that the director field had 4-fold rotational symmetry. The axis of mirror symmetry of a hyperbolic hedgehog became the domain boundary between the unit cells. Each hyperbolic hedgehog was shared by two unit cells (white solid lines in Fig. 6(a)). The saddle was shared by four neighboring unit cells. On the other hand, the hexagonal array had 6-fold rotational symmetry, since every 60° of rotation gave the same optical image. Every side of the hexagon (walls in 3D) has a half hyperbolic hedgehog embedded on the surface (white solid lines in Fig. 6(b)). Note that the saddle here is shared by three neighboring corners of three hexagons. The hyperbolic hedgehog between two unit cells splits the two neighboring unit cells.

Referring to the colors of the POM images, the director field of the topological defect network was estimated and illustrated in Fig. 7. A pad generates the radial hedgehog. A fishbone electrode simply emphasizes the diagonal components of the radial hedgehog. In a coil pixel, the directors circle around the defect, generating a spiral hedgehog. Circular hedgehogs were not observed. The director field inside the pixel are successfully controlled by the strips of the electrodes, but not affected by the lattice type (square or hexagonal).

Fig. 8 shows the topological defect array in a wide range. The scale of each picture is 1.7 mm × 1.2 mm. The square and fishbone pixels could form a regular, periodic, stable topological defect network (Fig. 8(a), (b), (d), and (e)). The size of defect network could be larger than 1 cm$^2$. In contrast, the spiral hedgehogs generated by coil pixels could not be arranged in a regular array. Most of the spiral hedgehogs deviated from the center of the pixel, or transformed into radial hedgehogs (Fig. 8(c) and (f)). Therefore, the distribution of director field was irregular. The spiral hedgehogs could easily drift away from the designed position, merged with the hyperbolic hedgehog nearby, and then the defects disappeared. The defect network composed of spiral hedgehogs was instable.

To summarize, arrays of point defects were successfully generated in vertically aligned liquid crystal cell when electric field was applied. All defects were points. Lines and disclinations had not been observed. The radial hedgehogs were positioned by design, and the hyperbolic hedgehogs where generated on the domain boundary spontaneously. Defects were arranged in square or hexagonal lattices like square or hexagonal crystals.

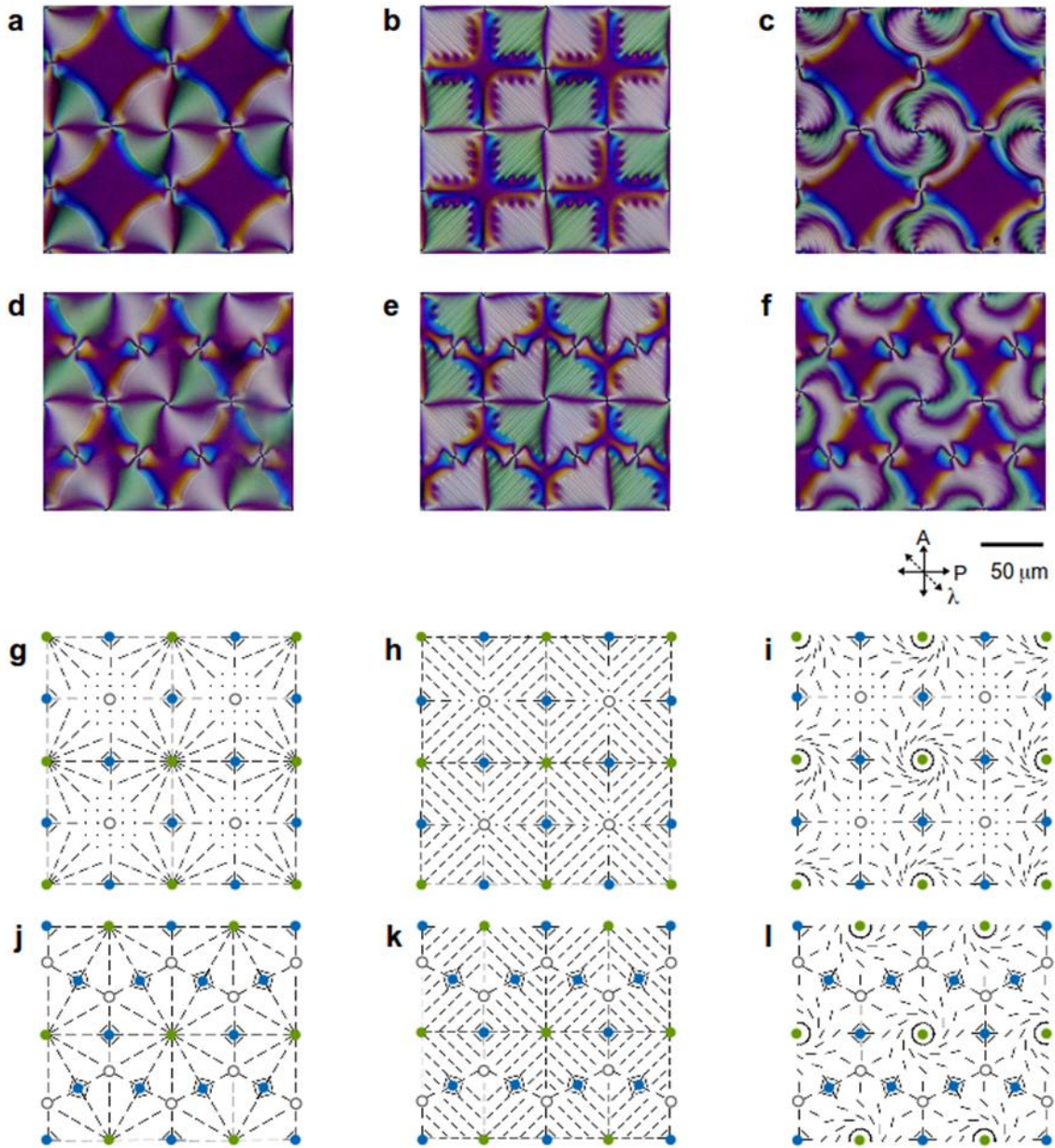

**Figure 7 Director field of defect arrays.** (a) - (f) Pictures of square and hexagonal arrays under polarized microscope. Crossed polarizers on 0° and 90°. λ-wave plate on 135°. (g) – (l) Corresponding director fields. (a) (g) Pad pixel in square lattice. (b) (h) Fishbone pixel in square lattice. (c) (i) Coil pixels in square lattice. (d) (j) Pad pixel in hexagonal lattice. (e) (k) Fishbone pixel in hexagonal lattice. (f) (l) Coil in hexagonal lattice. Green, blue, and white dots indicate the position of radial hedgehog, hyperbolic hedgehog, and saddle deformations, respectively. The black short lines show the director of liquid crystal.

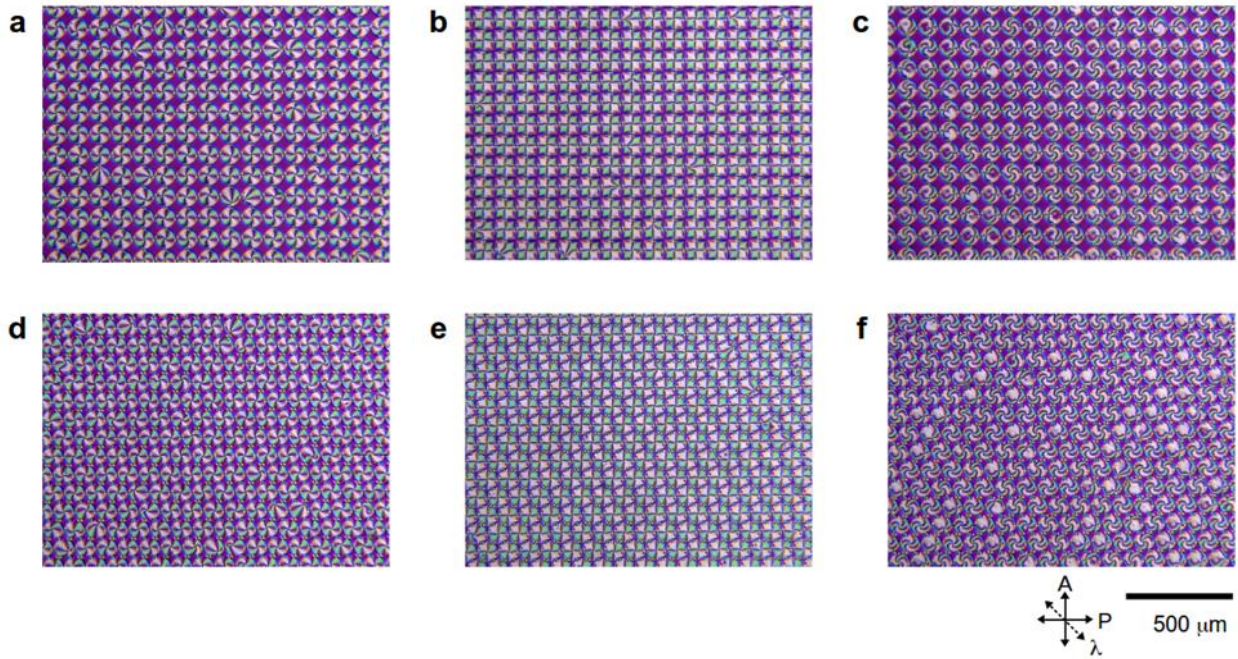

**Figure 8 Large range of defect arrays under polarized optical microscope.** Crossed polarizers on 0° and 90°. λ-wave plate on 135°. (a) Pad pixel in square lattice. (b) Fishbone pixel in square lattice. (c) Coil pixels in square lattice. (d) Pad pixel in hexagonal lattice. (e) Fishbone pixel in hexagonal lattice. (f) Coil in hexagonal lattice.

## Discussion

The radial (or spiral) hedgehogs, hyperbolic hedgehogs, and saddles were successfully arranged in two dimensional (2D) square and hexagonal lattices (Fig. 9 (a) and (c)). The arrays of topological defects were stable, and span over a large area. Though the information on the 3rd dimension (the z-axis) is limited under POM, the author does not want to be limited in two dimensions. The three dimensional (3D) unit cells can be deduced based on principles of symmetry and continuity of director field (Fig. 9 (b) and (d)). The director field on layer II in Fig. 9 was clearly identified in POM images (Fig. 7). Layer I and III are above and beneath Layer II, respectively. The defects on layer I and III can be derived according to the following principles:

1. A saddle is created at the middle of two out-going radial hedgehogs
2. A hyperbolic hedgehog is created at the middle of two saddles
3. The saddles and hyperbolic hedgehogs generate a boundary of which all the directors are pointing into the unit cell, resulting in the in-going radial hedgehog

The topological defects on layer I and III can be derived. Therefore the 3D cuboid and hexagonal cylinder unit cells are postulated, as illustrated in Fig. 9(b) and (d), respectively. The director field and topological charges of the 3D unit cells were analyzed to find the hidden conservation laws.

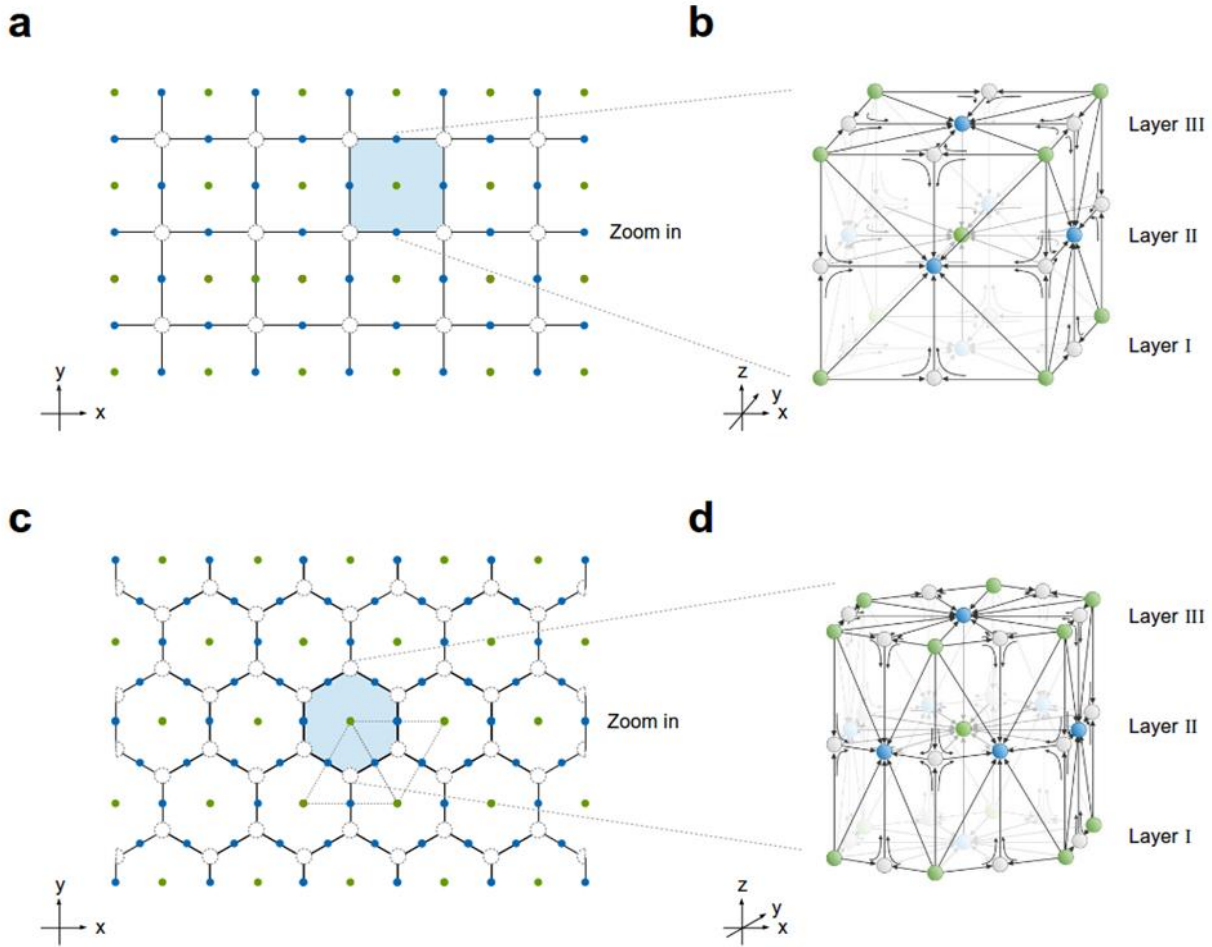

**Figure 9 Crystal of topological defects and unit cells.** (a) 2D square crystal (b) 3D cuboid unit cell (c) 2D hexagonal crystal (d) 3D hexagonal cylinder unit cell. The area in light blue are the unit cell of the crystal. Black lines show the domain boundary. Green, blue, and white dots indicate the position of radial hedgehog, hyperbolic hedgehog, and saddle deformations, respectively. The unit cell of a hexagonal crystal can be a hexagon or two anti-parallel triangles (marked by the red dashed lines). The black arrows show the director field.

The detailed director field of topological defects are illustrated in Fig. 10. The mathematical representation of director filed in spherical coordinates are summarized in Fig. 10, too. Note that the axis of rotational symmetry ($C_\infty$) of hyperbolic hedgehogs are normal to the surface of unit cells. The director field of a saddle is very similar to that of a hyperbolic hedgehog, but all the directors are reversed, which leads to a reversed hyperbolic hedgehog. The $C_\infty$ of saddles is along the edge of unit cells.

The liquid crystal molecules are considered as vectors here, though they have the head-to-tail symmetry. Due to the anchoring on the surface, the head-to-tail symmetry is broken. The end fixed on the surface is defined to be the tail. The other end that can move freely with respect to the applied electric field is defined as the head. The vectors are indicated by arrows.

**Conservation of Director Flux**

The director field created by topological defects seems very similar to the electric field created by electric charges. Though the electric field is a graphical representation of force and it does not move (but the force drives the charge moving), the electric field ($\vec{E}$) and charges ($q$) are still considered as flux and its sources (or sinks) in Electromagnetism, respectively. The relation between the flux and source is indicated in Gauss Law:

$$\oiint \vec{E} \cdot d\vec{A} = \frac{q}{\varepsilon}$$

where $d\vec{A}$ is the unit area indicated by a normal vector, $\varepsilon$ is dielectric constant, $q/\varepsilon$ represents the strength of the source (or the sink) in terms of electric charge ($q$).

Can we build a model for director field generated topological defects on the basis of the same terminology? The relation between the director flux and the topological defects are postulated in the following equation:

$$\oiint \vec{n} \cdot d\vec{A} = \text{strength of source or sink}$$

where the strength of the source (or sink) is proportional to the flux of directors from (or to) the topological defects.

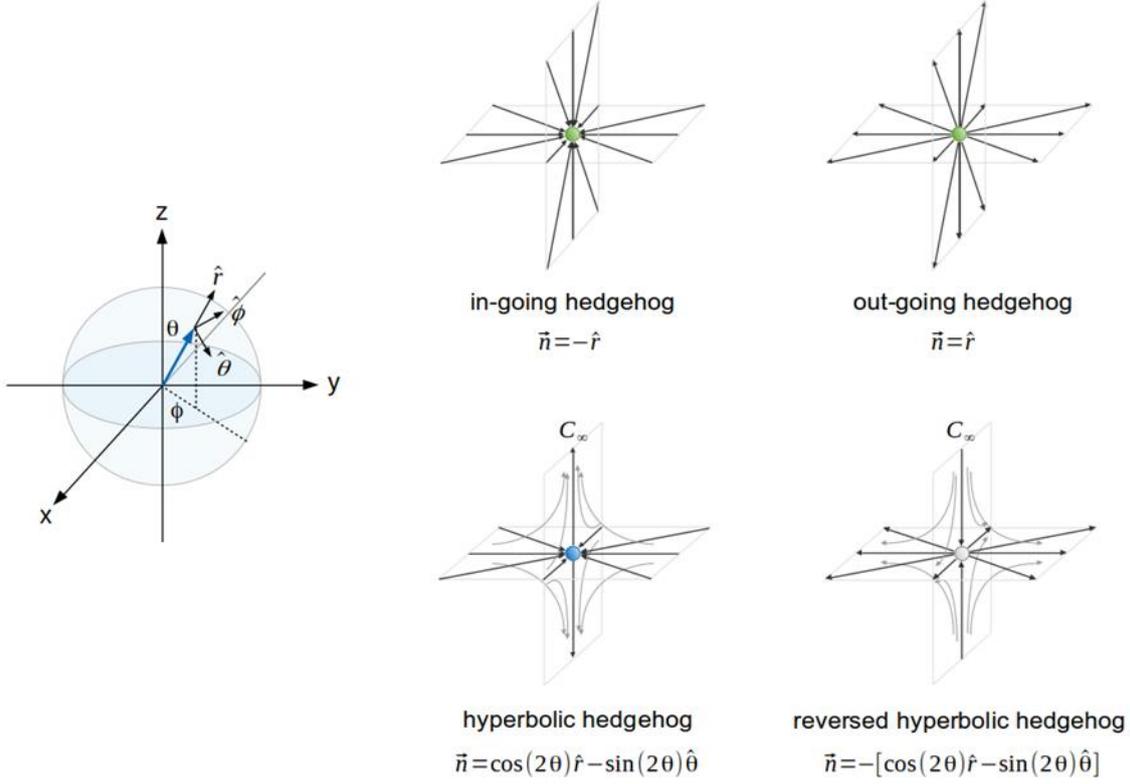

**Figure 10 Director field around point defects in details.** Black and grey arrows show the director field. $C_\infty$ marks the axis of rotational symmetry. The planes are the mirror of reflection symmetry. Spherical coordinate system on the left hand side. Mathematical representations of director field are under each schematic diagram.

**Flux in Cuboid Crystal** The flux of directors from (or into) a topological defect is calculated below. The director field of radial hedgehogs are described by the following expressions in spherical coordinates (see the coordinate system in Fig. 10):

$$\text{Out-going radial hedgehog:} \quad \vec{n} = \hat{r}$$
$$\text{In-going radial hedgehog:} \quad \vec{n} = -\hat{r}$$

where $\vec{n}$ is the director and $\hat{r}$ is the unit vector on the radial direction. The flux of directors from a radial hedgehog ($\Phi_{hr+}$) is the integration of $\vec{n} \cdot d\vec{A}$ over a closed surface that wraps the hedgehog up:

$$\Phi_{hr+} = \oiint \vec{n} \cdot d\vec{A} = \oiint \hat{r} \cdot d\vec{A} = \oiint \hat{r} \cdot (r d\theta \hat{\theta} \times r \sin\theta \, d\phi \hat{\phi})$$
$$= r^2 \int_0^\pi \sin\theta \, d\theta \int_0^{2\pi} d\phi = 4\pi r^2$$

The flux of directors flowing into a radial hedgehog ($\Phi_{hr-}$) is

$$\Phi_{hr-} = \oiint \vec{n} \cdot d\vec{A} = \oiint -\hat{r} \cdot d\vec{A} = -4\pi r^2$$

Because $4\pi r^2$ is the surface area of the sphere, which scales with the radius r, the strength of radial hedgehog as a source (or sink) can be defined by the flux per unit area, namely, the "director flux charge". The director flux charge of an out-going radial hedgehog ($Q_{hr+}$) and an in-going radial hedgehog ($Q_{hr-}$) are given by the calculation below:

$$Q_{hr+} = \frac{\Phi_{hr+}}{4\pi r^2} = +1$$

$$Q_{hr-} = \frac{\Phi_{hr-}}{4\pi r^2} = -1$$

The director field of a hyperbolic hedgehog is given by the equation:

$$\text{hyperbolic hedgehog: } \vec{n} = \left[\cos(2\theta)\hat{r} - \sin(2\theta)\hat{\theta}\right]$$

Here the hyperbolic hedgehog has the rotational symmetry about the z-axis, and $\theta$ is the angle between $\vec{n}$ and the axis of rotational symmetry. The reversed hyperbolic hedgehog is obtained by reversing the directors of a hyperbolic hedgehog:

$$\text{Reversed hyperbolic hedgehog: } \vec{n} = -\left[\cos(2\theta)\hat{r} - \sin(2\theta)\hat{\theta}\right]$$

Saddle deformation and reversed hyperbolic hedgehog have the same director field, though saddle does not have a defect core. The flux of directors from a hyperbolic hedgehog ($\Phi_{hh+}$) is:

$$\Phi_{hh+} = \oiint \vec{n} \cdot d\vec{A} = \oiint \left[\cos(2\theta)\hat{r} - \sin(2\theta)\hat{\theta}\right] \cdot \left(rd\theta\hat{\theta} \times r\sin\theta\, d\phi\hat{\phi}\right)$$

$$= r^2 \int_0^\pi \cos(2\theta)\sin\theta\, d\theta \int_0^{2\pi} d\phi = r^2 \int_0^\pi (2\cos^2\theta - 1)(-d\cos\theta) \int_0^{2\pi} d\phi$$

$$= -\frac{4}{3}\pi r^2$$

The flux from a reversed hyperbolic hedgehog ($\Phi_{hh-}$) is

$$\Phi_{hh-} = \oiint \vec{n} \cdot d\vec{A} = \oiint -\left[\cos(2\theta)\hat{r} - \sin(2\theta)\hat{\theta}\right] \cdot \left(rd\theta\hat{\theta} \times r\sin\theta\, d\phi\hat{\phi}\right) = \frac{4}{3}\pi r^2$$

Then the director charge of a hyperbolic hedgehog ($Q_{hh+}$) and a reversed hyperbolic hedgehog ($Q_{hh-}$) are given as below:

$$Q_{hh+} = \frac{\Phi_{hh+}}{4\pi r^2} = -\frac{1}{3}$$

$$Q_{hh-} = \frac{\Phi_{hh-}}{4\pi r^2} = +\frac{1}{3}$$

The flux charge of all types of 3D point defects are summarized in Table 1.

As illustrated in Fig. 9(b), directors are from the out-going radial hedgehogs on the 8 corners of the unit cell. They turn the direction at the hyperbolics and saddles, and finally gather at the in-going radial hedgehog at the center of the cell. The structure of layer I and III are the same. The directors on the walls were in plane, but they switch to the direction normal to the wall via the saddles and hyperbolic hedgehogs on Layer II. Finally, all the directors "flow" into the center of the unit cell, and the in-going radial hedgehog is generated.

On layer I and III, an out-going radial hedgehog is shared by 8 neighboring unit cells, one saddle is shared by 4 neighboring unit cells, and the hyperbolic hedgehog at the center of a wall is shared by 2 neighboring unit cells. On layer II, an in-going hedgehog is entirely in the unit cell, a saddle on the corner is shared by 4 neighboring cells, and a hyperbolic hedgehog on the side is shared by 2 neighboring unit cells. The flux charges on each layer are indicated in Fig. 11(a) and summarized in the following equations:

                    **Center**                     **Corner**               **Sides**

**Layer I**:  $1 \times \left[\frac{1}{2} \text{hyperbolic hedgehog}\right] + 4 \times \left[\frac{1}{8} \text{out-going radial hedgehog}\right] + 4 \times \left[\frac{1}{4} saddle\right]$

**Layer II**:  $1 \times [1 \text{ in-going radial hedgehog}] + 4 \times \left[\frac{1}{4} saddle\right] + 4 \times \left[\frac{1}{2} \text{hyperbolic hedgehog}\right]$

**Layer III**:  $1 \times \left[\frac{1}{2} \text{hyperbolic hedgehog}\right] + 4 \times \left[\frac{1}{8} \text{out-going radial hedgehog}\right] + 4 \times \left[\frac{1}{4} saddle\right]$

Therefore:

                    **Center**         **Corner**        **Sides**

**Layer I**:       $1 \times \left[\frac{1}{2} Q_{hh+}\right] + 4 \times \left[\frac{1}{8} Q_{hr+}\right] + 4 \times \left[\frac{1}{4} Q_{hh-}\right]$

**Layer II**:      $1 \times [1\, Q_{hr-}] + 4 \times \left[\frac{1}{4} Q_{hh-}\right] + 4 \times \left[\frac{1}{2} Q_{hh+}\right]$

**Layer III**:     $1 \times \left[\frac{1}{2} Q_{hh+}\right] + 4 \times \left[\frac{1}{8} Q_{hr+}\right] + 4 \times \left[\frac{1}{4} Q_{hh-}\right]$

                    **Center**         **Corner**        **Sides**

**Layer I**:    $Q_I = 1 \times \left[\frac{1}{2}\left(-\frac{1}{3}\right)\right] + 4 \times \left[\frac{1}{8}(+1)\right] + 4 \times \left[\frac{1}{4}\left(+\frac{1}{3}\right)\right] = -\frac{1}{6} + \frac{1}{2} + \frac{1}{3}$

**Layer II**:   $Q_{II} = 1 \times [1(-1)] + 4 \times \left[\frac{1}{4}\left(+\frac{1}{3}\right)\right] + 4 \times \left[\frac{1}{2}\left(-\frac{1}{3}\right)\right] = -1 + \frac{1}{3} - \frac{2}{3}$

**Layer III**:  $Q_{III} = 1 \times \left[\frac{1}{2}\left(-\frac{1}{3}\right)\right] + 4 \times \left[\frac{1}{8}(+1)\right] + 4 \times \left[\frac{1}{4}\left(+\frac{1}{3}\right)\right] = -\frac{1}{6} + \frac{1}{2} + \frac{1}{3}$

The total "flux charge" in a lattice is

$$Q_I + Q_{II} + Q_{III} = 0$$

which reveals that the total director flux in a cuboid unit cell is 0, indicating that the directors pointing into the unit cell must equal to the directors gathered at the center defect. The director flux is conserved in a stable cuboid unit cell. Since crystal is a periodic repetition of unit cells, the flux is conserved in the whole defect crystal, too.

    The cuboid unit cell with a radial hedgehog at the center, created by the pad and fishbone electrode, which leads to stable and robust crystal of defects, is the positive case for flux conservation. On the other hand, the circular hedgehog at a cuboid unit cell, created by coil electrode, is instable in the defect array, which is the negative example for flux conservation. The center defect turns into a spiral hedgehog (a mixture of radial and circular components), or it simply shows up as a radial hedgehog of which the location is out of control.

    The director field of a circular hedgehog is

$$\vec{n} = \hat{\phi}$$

and the flux of a circular hedgehog ($\Phi_{hc}$) is

$$\Phi_{hc} = \oiint \vec{n} \cdot d\vec{A} = \oiint \hat{\phi} \cdot (rd\theta\hat{\theta} \times r\sin\theta\, d\phi\hat{\phi}) = 0$$

The flux of directors is a closed loop in the circular defect, so the flux is zero. Thus the flux charge ($Q_{hc}$) is 0, too:

$$Q_{hc} = 0$$

In this case, the total director flux in a unit cell is not zero:

$$Q_I + Q_{II} + Q_{III} = +1 \neq 0$$

indicating that the directors pointing into the unit cell is more than the directors absorbed by the center defect. The flux from the source is not consistent with the flux to the sink, and therefore the crystal of defect is not stable. The directors from the source rush into unexpected directions to release the flux to the sink. Hence the crystal of circular hedgehogs collapse in periodic homeotropic boundaries.

**Flux in Hexagonal Cylinder Crystal** In a hexagonal cylinder unit cell, the directors are from twelve out-going radial hedgehogs on layer I and layer III. The planar directors on the walls switch into a vertical alignment via the hyperbolic hedgehogs and saddles. Finally, all the directors point into the center of the unit cell, generating an in-going radial hedgehog. Actually, the unit cells in a hexagonal crystal must be triangles arranged in anti-parallel manner, as indications in Fig. 9(c). Each corner of the triangular unit cell is 1/6 of an in-going radial hedgehog. Therefore, a saddle contributes only 3×(1/6) of director flux on Layer II. Here, we keep using the hexagonal cylinder unit cell to have consistent analysis on the data and in the models. Distribution of hyperbolic hedgehogs and saddles on the walls are the same as the cases in cuboid unit cells. The three layers of defect network are indicated in Fig. 11(b) and summarized the following equations:

$$
\begin{array}{ll}
& \quad\quad\quad\quad\text{Center} \quad\quad\quad\quad\quad\quad\quad\quad \text{Corner} \quad\quad\quad\quad\quad\quad\quad \text{Sides} \\
\textbf{Layer I}: & 1 \times \left[\tfrac{1}{2}\text{hyperbolic hedgehog}\right] + 6 \times \left[\tfrac{1}{6}\cdot\tfrac{1}{2}\text{out-going radial hedgehog}\right] + 6 \times \left[\tfrac{1}{4} saddle\right] \\
\textbf{Layer II}: & 1 \times [1 \text{ in-going radial hedgehog}] + 6 \times \left[\tfrac{1}{3}\cdot\tfrac{1}{2}\text{saddle}\right] + 6 \times \left[\tfrac{1}{2}\text{hyperbolic hedgehog}\right] \\
\textbf{Layer III}: & 1 \times \left[\tfrac{1}{2}\text{hyperbolic hedgehog}\right] + 6 \times \left[\tfrac{1}{6}\cdot\tfrac{1}{2}\text{out-going radial hedgehog}\right] + 6 \times \left[\tfrac{1}{4} saddle\right]
\end{array}
$$

Then the flux charges on each layer are

$$
\begin{array}{ll}
& \quad\quad\quad \text{Center} \quad\quad\quad \text{Corner} \quad\quad \text{Sides} \\
\textbf{Layer I}: & 1 \times \left[\tfrac{1}{2}Q_{hh+}\right] + 6 \times \left[\tfrac{1}{12}Q_{hr+}\right] + 6 \times \left[\tfrac{1}{4}Q_{hh-}\right] \\
\textbf{Layer II}: & 1 \times [1\, Q_{hr-}] + 6 \times \left[\tfrac{1}{6}Q_{hh-}\right] + 6 \times \left[\tfrac{1}{2}Q_{hh+}\right] \\
\textbf{Layer III}: & 1 \times \left[\tfrac{1}{2}Q_{hh+}\right] + 6 \times \left[\tfrac{1}{12}Q_{hr+}\right] + 6 \times \left[\tfrac{1}{4}Q_{hh-}\right]
\end{array}
$$

$$
\begin{array}{ll}
& \quad\quad\quad\quad \text{Center} \quad\quad\quad\quad \text{Corner} \quad\quad\quad \text{Sides} \\
\textbf{Layer I}: & Q_I = 1 \times \left[\tfrac{1}{2}\left(-\tfrac{1}{3}\right)\right] + 6 \times \left[\tfrac{1}{12}(+1)\right] + 6 \times \left[\tfrac{1}{4}\left(+\tfrac{1}{3}\right)\right] = -\tfrac{1}{6} + \tfrac{1}{2} + \tfrac{1}{2} \\
\textbf{Layer II}: & Q_{II} = 1 \times [1(-1)] + 6 \times \left[\tfrac{1}{6}\left(+\tfrac{1}{3}\right)\right] + 6 \times \left[\tfrac{1}{2}\left(-\tfrac{1}{3}\right)\right] = -1 + \tfrac{1}{3} - 1 \\
\textbf{Layer III}: & Q_{III} = 1 \times \left[\tfrac{1}{2}\left(-\tfrac{1}{3}\right)\right] + 6 \times \left[\tfrac{1}{12}(+1)\right] + 6 \times \left[\tfrac{1}{4}\left(+\tfrac{1}{3}\right)\right] = -\tfrac{1}{6} + \tfrac{1}{2} + \tfrac{1}{2}
\end{array}
$$

The total "director charge" in a lattice is

$$Q_I + Q_{II} + Q_{III} = 0$$

The total flux is 0 again. The flux from the source is equal to the flux to the sink. The director flux is conserved in hexagonal unit cell, too. However, there is more radial component in a hexagonal unit cell than in a cuboid one, since hexagonal unit cells have more saddles and hyperbolic hedgehogs on the corners and edges. Therefore, radial and spiral hedgehogs are stabilized, and circular hedgehogs are not favored in hexagonal crystals.

**Conservation of Topological Charges**

The summation of topological charges in confined liquid crystal is determined by the Euler characteristic ($\chi$) of the confinement. To have a clear logic flow in the following derivation, topological charges are classified into surface charges ($s$) and bulk charges ($m_b$). For a closed confinement with planar surface[40,41], the summation of surface charges ($s_i$) is

$$\sum_i s_i = \chi$$

where $i$ is the index. The equation was derived from Poincaré-Hopf theorem. For homeotropic surfaces[3,4], the total bulk charges in the confinement is

$$\sum_k m_{b,k} = \frac{\chi}{2}$$

where k is the index. The relation is from the Gauss-Stein Theorem for characteristic of surfaces and Gauss-Bonnet Theorem for singularities in vector field. The sum of surface charge on a homeotropic confinement must be 0[42]. The Euler characteristic is topologically invariant. Though the number of charges can change due to phase transition, Fréedericksz transition, or variation of geometries, the summation of topological charges must conserve. $\chi$ of a closed sphere is 2. A cuboid or hexagonal cylinder surface is topologically equivalent to a sphere, so the $\chi$ of our unit cell is always 2. A single unit cell with homeotropic surface obeys the equation, $\sum_k m_{b,k} = \chi/2$. However, additional defects are created on the boundary between unit cells in the lattice. If topological charge is conserved in a closed confinement, the total $s$ and $m_b$ must be 0 and 1[42], respectively, for both cuboid and hexagonal cylinder unit cells. Additionally, the $m_b$ inside the confinement and the summation of $m_b$ embedded on the boundary must equal to each other, because the degree of continuous mapping must conserve[3], too.

The bulk charge is the wrapping number of closed surfaces surrounding a defect, which can be calculated by the following equation[43]:

$$m_b = \frac{1}{4\pi} \oiint \left[ \vec{n} \cdot \left( \frac{\partial \vec{n}}{\partial \theta} \times \frac{\partial \vec{n}}{\partial \phi} \right) \right] d\theta d\phi$$

The projection of 3D director field around a bulk defect onto the surface of its confinement is the 2D surface defect, and its strength is the winding number of closed loops surrounding the defect core:

$$s = \frac{1}{2\pi} \oint \left( \vec{n} \cdot \frac{\partial \vec{n}}{\partial \phi} \right) \cdot d\vec{l}$$

The $m_b$ and $s$ of radial hedgehogs are always +1. The $m_b$ of regular and reversed hyperbolic hedgehogs are +1 and -1[44], respectively. The $s$ of hyperbolic hedgehogs depends on the surface for projection. As shown in Fig. 10, when the $C_\infty$ axis of a hyperbolic hedgehog is the z-axis, its $s$ is +1 on the x-y plane, but its projection on y-z plane results in $s$ equal to -1[44,45]. The $s$ of a reversed hyperbolic hedgehog is obtained in the same manner. The topological charges of all the hedgehogs are summarized in Table 1.

The positions of the topological defects are illustrated in Fig. 11(c) and (d) for 3D cuboid and hexagonal cylinder unit cells, respectively. The $m_b$ and $s$ are marked next to the defects. In a cuboid unit cell, the $m_b$ of radial hedgehog inside the cell is +1. The $m_b$ of defects embedded on the surfaces are summarized in the equations below:

**Center  Corner  Sides**

**Layer I**: $m_{b,I} = 1 \times \left[\frac{1}{2}(+1)\right] + 4 \times \left[\frac{1}{8}(+1)\right] + 4 \times \left[\frac{1}{4}(-1)\right] = +\frac{1}{2} + \frac{1}{2} - 1$

**Layer II**: $m_{b,II} = 0 \times [0(+0)] + 4 \times \left[\frac{1}{4}(-1)\right] + 4 \times \left[\frac{1}{2}(+1)\right] = +0 - 1 + 2$

**Layer III**: $m_{b,III} = 1 \times \left[\frac{1}{2}(+1)\right] + 4 \times \left[\frac{1}{8}(+1)\right] + 4 \times \left[\frac{1}{4}(-1)\right] = +\frac{1}{2} + \frac{1}{2} - 1$

The total $m_b$ of defects embeded on the surface is then

$$m_{b,I} + m_{b,II} + m_{b,III} = +1$$

The bulk topological charge obeys the Gauss-Stein Theorem for homeotropic boundary. Then the $m_b$ of center bulk defect is equivalent to the total $m_b$ of surface defects because the degree of mapping conserves.

The surface charge, $s$, of each wall of the unit cell is summarized in the following equation:

**Center  Corner  Sides**

**ceiling**: $s_{ceiling} = 1 \times [+1] + 4 \times \left[+\frac{1}{4}\right] + 4 \times \left[-\frac{1}{2}\right] = +1 + 1 - 2$

**wall**: $s_{wall} = 1 \times [+1] + 4 \times \left[+\frac{1}{4}\right] + 4 \times \left[-\frac{1}{2}\right] = +1 + 1 - 2$

**floor**: $s_{floor} = 1 \times [+1] + 4 \times \left[+\frac{1}{4}\right] + 4 \times \left[-\frac{1}{2}\right] = +1 + 1 - 2$

The total $s$ of defects projected on the boundary is then

$$s_{ceiling} + 4 \cdot s_{wall} + s_{floor} = 0$$

which is identical to the total $s$ of a homeotropic surface.

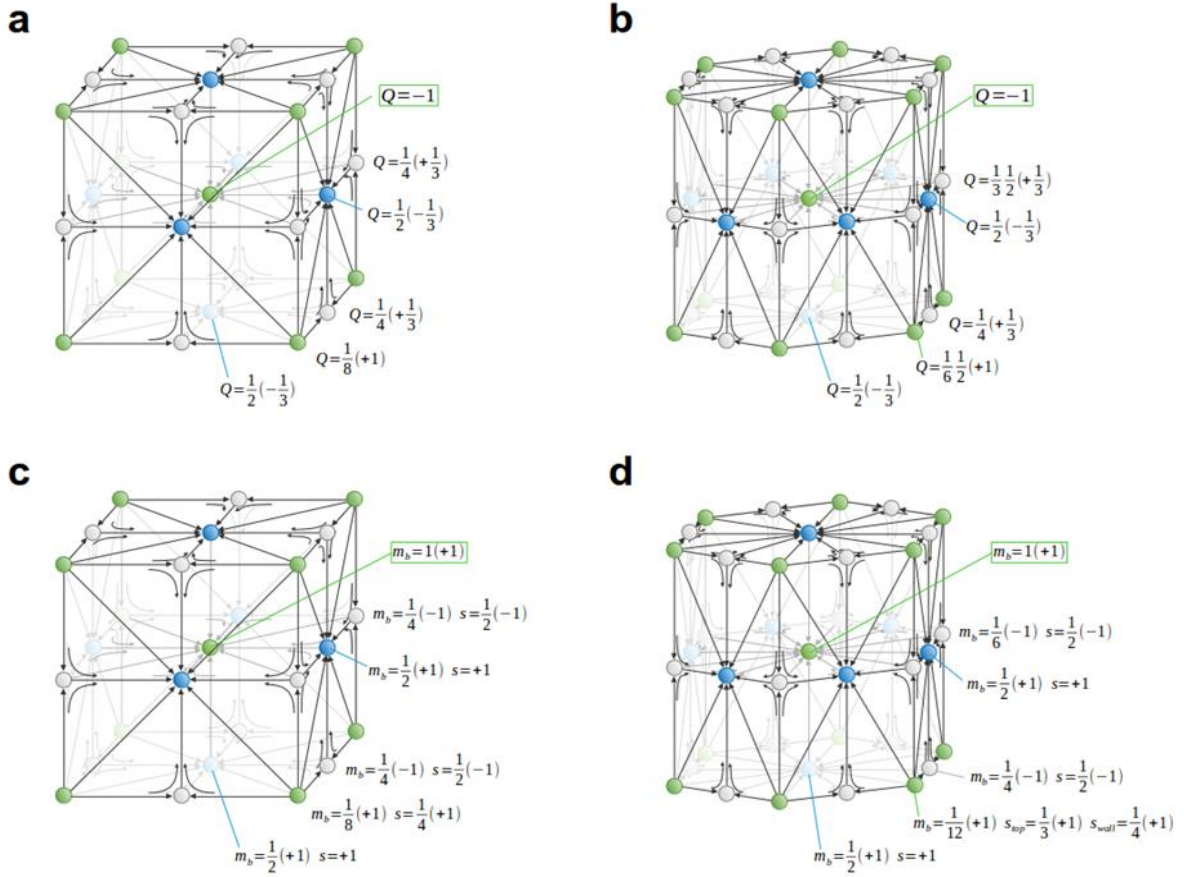

**Figure 10 Charges of unit cells.** (a) Flux charge of director flux in a cuboid unit cell. (b) Flux charge in a hexagonal unit cell. (c) Topological charge in a cuboid unit cell. (d) Topological charge in a hexagonal unit cell. $Q$ is flux charge. $m_b$ and $s$ are bulk and surface topological charge, respectively. Green, blue, and white dots indicate the position of radial hedgehog, hyperbolic hedgehog, and saddle deformations, respectively. The black arrows show the director field.

The same analysis is performed for the hexagonal crystal. In a hexagonal cylinder unit cell, the $m_b$ in the bulk is +1, too. The $m_b$ of defects embedded on the surfaces are summarized in the following equations:

$$\begin{array}{cccc} & \text{Center} & \text{Corner} & \text{Sides} \\ \textbf{Layer I}: & m_{b,I} = 1 \times \left[\frac{1}{2}(+1)\right] + 6 \times \left[\frac{1}{12}(+1)\right] + 6 \times \left[\frac{1}{4}(-1)\right] = +\frac{1}{2} + \frac{1}{2} - \frac{3}{2} \\ \textbf{Layer II}: & m_{b,II} = 0 \times [0(+0)] + 6 \times \left[\frac{1}{6}(-1)\right] + 6 \times \left[\frac{1}{2}(+1)\right] = +0 - 1 + 3 \\ \textbf{Layer III}: & m_{b,III} = 1 \times \left[\frac{1}{2}(+1)\right] + 6 \times \left[\frac{1}{12}(+1)\right] + 6 \times \left[\frac{1}{4}(-1)\right] = +\frac{1}{2} + \frac{1}{2} - \frac{3}{2} \end{array}$$

$$m_{b,I} + m_{b,II} + m_{b,III} = +1$$

For the surface defects:

$$\begin{array}{cccc} & \text{Center} & \text{Corner} & \text{Sides} \\ \textbf{ceiling}: & s_{ceiling} = 1 \times [+1] + 6 \times \left[+\frac{1}{3}\right] + 6 \times \left[-\frac{1}{2}\right] = +1 + 2 - 3 \\ \textbf{wall}: & s_{wall} = 1 \times [+1] + 4 \times \left[+\frac{1}{4}\right] + 4 \times \left[-\frac{1}{2}\right] = +1 + 1 - 2 \\ \textbf{floor}: & s_{floor} = 1 \times [+1] + 6 \times \left[+\frac{1}{3}\right] + 6 \times \left[-\frac{1}{2}\right] = +1 + 2 - 3 \end{array}$$

$$s_{ceiling} + 6 \cdot s_{wall} + s_{floor} = 0$$

In both cuboid and hexagonal cylinder unit cells,

$$\sum_{bulk} m_b = \sum_{surface} m_b = \frac{\chi}{2}$$

$$\sum_{surface} s = 0$$

The bulk topological charge is determined by the Euler characteristic of the confinement. If the shapes of confinements are topologically invariant, the total $m_b$ conserves. The surface charge is 0 for homeotropic, closed surface. It is independent of the shape of confinement, and conserve under continuous morphing, too.

Note that the radial and circular hedgehog has the identical $m_b$ and $s$, which are equal to 1. If a radial hedgehog were replaced by a circular one at the center of unit cells, the total $m_b$ and $s$ would be unchanged. However, the experimental results show that circular hedgehog is instable in crystal of defects. The difference between radial and circular hedgehogs is well explained in terms of conservation flux: The total flux is 0 if the bulk defect is a radial hedgehog, while the flux from the source and the flux to the sink cannot be equal if the bulk defect is a circular hedgehog. Radial and circular hedgehogs are identical in the analysis of topological charges, so the conservation of topological charges is not capable to predict the stability of defect network in this case.

**Summary**

The charges for conservation laws are summarized in Table 1. For crystal of defects generated in homeotropic, period confinement, conservation laws are summarized in the following equations:

$$\sum_{bulk} Q = 0$$

$$\sum_{bulk} m_b = \frac{\chi}{2}$$

$$\sum_{surface} s = 0$$

The flux (represented by charge, $Q$), bulk topological charge ($m_b$), and surface topological charge ($s$) conserve under continuous morphing of the unit cells, and must be invariant under deformations

**Physical meaning of flux charge conservation** Flux charge represents the divergence of director field round the defect. Divergence of the director field is a measurement for splay deformation, which is dominant in liquid crystal confined in homeotropic surfaces. The elastic potential energy is proportional to the square of the divergence[46], and the gradient of the divergence is the force between defects in a defect crystal. The self-organizing behavior of particles in liquid crystal is driven by this kind of long-range force[10]. The conservation of flux charge implies that the long-range force is balanced and the potential energy is minimized. The crystal of topological defects mist be a ground state of energy and thus stabilized.

**Physical meaning of topological charge conservation** The conservation of topological charge is determined and protected by the topological structure of the confinement. It usually applies to soft, self-retained, self-healing structure because the topological charges are invariant under continuous distortion. Topological charge conservation also implies that the sufficient combination of defects in a stable defect crystal should be quantized.

Table 1 Flux, bulk, and surface charges of hedgehogs. The coordinate system is in Fig. 10.

|  | Out-going radial hedgehog (prefix: hr+) | In-going radial hedgehog (prefix: hr-) | Hyperbolic hedgehog (prefix: hh+) | Reversed hyperbolic hedgehog (prefix: hh-) |
|---|---|---|---|---|
| Flux charge ($Q$) | +1 | +1 | $-\frac{1}{3}$ | $+\frac{1}{3}$ |
| Bulk topological charge ($m_b$) | +1 | +1 | +1 | -1 |
| Surface topological charge ($s$) on x-y plane | +1 | +1 | +1 | +1 |
| Surface topological charge ($s$) on y-z plane | +1 | +1 | -1 | -1 |

# Conclusion

Analysis on the director flux and topological charges reveals the conservation laws for a stable defect array. The conservation laws may explain why certain long-range order exist in nature, while some of the structure collapse because of instability. They can be implemented in templates for self-assembled micro-structure and tissue design. Calculation and minimization of free energy may lead to dynamics of self-assembled long-range order. The observation and discussion is limited in static crystal in this research. If the flow of directors or spinning of the defects is considered as flow and swirl in fluid, the conservation law may extend for dynamic and active behavior of soft matter, in analogy to continuity theorem of fluid dynamics. In this research, 2D defect crystal was generated, and the structure of 3D unit cell was deduced. By invention of periodic confinement in 3D space, hopefully inspired by the growth and forms of natural crystals and living tissues, 3D topological defect crystal is anticipated in the future.